\documentclass[twocolumn,prb, showpacs]{revtex4-1}
\usepackage{graphicx}
\usepackage{dcolumn}
\usepackage{bm}
\usepackage{color}

\begin{document}


\title{Spontaneous parity symmetry breaking of $\pi$-bonds in 2D:\\ describing topological insulators in real space}

\author{G. J. Shu$^1$}
\author{F. C. Chou$^{1,2,3}$}
\email{fcchou@ntu.edu.tw}
\affiliation{
$^1$Center for Condensed Matter Sciences, National Taiwan University, Taipei 10617, Taiwan}
\affiliation{
$^2$National Synchrotron Radiation Research Center, Hsinchu 30076, Taiwan}
\affiliation{
$^3$Center for Emerging Material and Advanced Devices, National Taiwan University, Taipei 10617, Taiwan}

\date{\today}

\begin{abstract}
The existence of $\pi$-bonds on topological insulator surfaces is found to be closely related to the phenomena of surface conduction and surface band spin polarization.  A $\pi$-bond trimer or $\pi$-bond dimer on the surface can form open conjugated systems that are responsible for the unique surface conduction mechanism of a topological insulator.  Parity operation in 2D is identified for a $\pi$-bond trimer within a six-fold symmetry coordination and a $\pi$-bond dimer within a four-fold symmetry coordination.  Spontaneous 2D parity symmetry breaking is found to be closely related to the theoretically predicted $\pi$ Berry's phase and the observed surface band spin polarization.  The role of $\pi$-bonds on a cleaved Bi$_2$Se$_3$ surface is compared to that for graphene with a  2D band structure containing Dirac cones.  Similar $\pi$-bond dimers within a four-fold 2D symmetry coordination can also be identified in strained $\alpha$-Sn  of diamond structure as a theoretically predicted topological insulator.       
\end{abstract}


\maketitle

\section{\label{sec:level1} introduction\protect\\ }

Topological insulators belong to a novel class of materials that exhibits band inversion characteristics due to strong spin-orbit interactions, which correspond to a unique state of matter for a band insulator with surface-only conduction and a spin-polarized surface state.  Bi$_2$Se$_3$ is a second-generation topological insulator with a nearly idealized single Dirac cone at the $\bar{\Gamma}$-point.\cite{Hasan2010}   The crystal structure of Bi$_2$Se$_3$ can be described in terms of close packed Bi-Se quintuple layers with a space group of R$\bar{3}$m and van der Waals gaps between the Se1-layers.\cite{Huang2012}  Topological insulator have long been modeled using analogous descriptions of topological invariants, Berry's phase, and the quantum spin Hall effect.\cite{Hasan2010, Qi2011}  In particular, the defining property of a topological insulator with a predicted chiral edge mode, i.e., the helical Dirac fermions of Z$_2$ topological-order has been confirmed through spin- and angle-resolved photoemission spectroscopy (spin-ARPES).\cite{Hsieh2009}   The success of many sophisticated theoretical predictions and subsequent experimental verifications since 2006 has led to rapid development in the research conducted on this exotic phase of electrons in solids.\cite{Kane2006, Hasan2010}  

Electricity conduction by topological insulator surfaces is mainly attributed to the band inversion phenomenon that has been proposed for heavy elements with strong spin-orbit coupling.\cite{Zhang2009}  Band inversion for the partially filled $p$ orbitals in Bi and Se may exist under crystal field splitting and strong spin-orbit coupling.  The crystal structure of Bi$_2$Se$_3$ can best be described in terms of a Bi-Se quintuple-layer unit, where the quintuple layers are close-packed with a R$\overline{3}$m space group symmetry in either a rhombohedral or hexagonal setting.\cite{Zhang2009, Huang2012}  Because Bi$_2$Se$_3$ has a band gap size of $\sim$0.3 eV in the semiconductor range and both Bi/Se elements are close-packed with octahedral coordination, we submit that Bi$_2$Se$_3$ may be adequately described by a fully hybridized $s$-$p$-$d$ orbital model, similar to $sp^3$ hybridization for Si with $3s^2p^2$, rather than by only partially filled $p$ orbitals within an ionic model in a point charge crystal field.  Recently, we proposed that the existence of $\pi$-bonds and Dirac cones in both graphene and the topological insulator surface of Bi$_2$Se$_3$ could be the key to explain the unique surface conduction mechanism in terms of a local $\pi$-bond energy exchange for graphene and  topological insulators.\cite{Shu2012}  In this study, we report 2D parity symmetry for $\pi$-bonds in topological insulators: the spontaneous 2D parity symmetry breaking may explain experimental observations of a spin-polarized gapless Dirac cone surface state that are consistent with theoretical predictions from band parity analysis.  
 
\section{\label{sec:level1} $\pi$-bonds on a $Bi_2Se_3$ surface\protect\\ }
   
\begin{figure}
\includegraphics[width=3.5in]{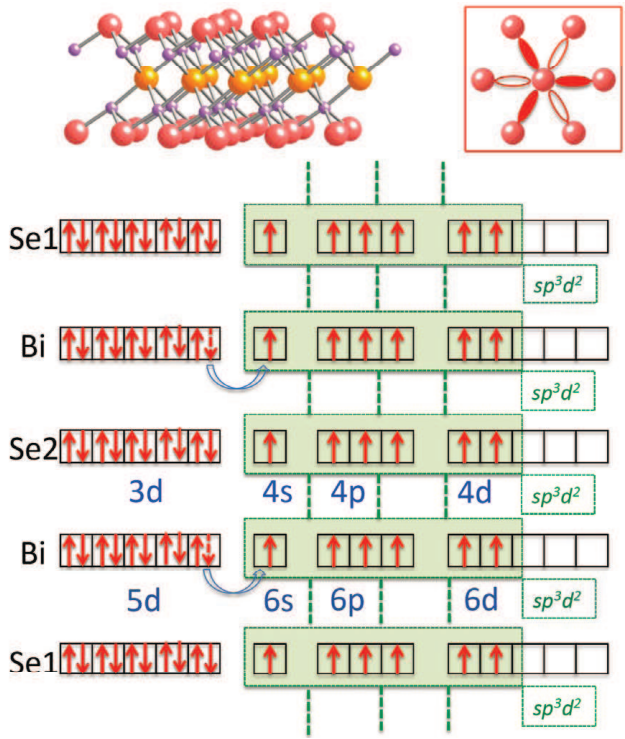}
\caption{\label{fig:fig-hybridBi2Se3}(color online) The proposed orbital hybridization for a quintuple Bi$_2$Se$_3$ layer with the actual crystal structure shown on the top left ( Se1 (red), Bi (purple) and Se2 (orange)) and the Se1 surface shown on the top right. The green dashed lines represent the six hybridized $sp^3d^2$ channels with an octahedral shape, in which $\sigma$-bonds are formed between the Bi and Se atoms, but the three unpaired $\pi$-electrons of Se1 at the van der Waals gap or surface may form a $\pi$-bond trimer with three of the six Se1 atoms in the same layer.}  
\vspace{-5mm}
\end{figure}

We first briefly review the hybrid orbital model for Bi$_2$Se$_3$ and the existence of $\pi$-bonds on the (111)$_r$ surface using the primitive rhombohedral axes (i.e., the (001)$_h$ in the hexagonal axes).\cite{Shu2012}  Considering the number of outer-shell electrons of Bi([Xe]4f$^{14}$5d$^{10}$6s$^2$6p$^3$) and Se([Ar]3d$^{10}$4s$^2$4p$^4$) and the close-packed BiSe$_6$-SeBi$_6$ octahedra, chemical bonding can be described by six hybridized $sp^3d^2$ orbitals to satisfy the correct coordination numbers with two electrons per $\sigma$-bond, as shown in Fig.~\ref{fig:fig-hybridBi2Se3}.  Both the Bi and Se2 atoms (in the central Se layer of the quintuple layers) can participate in the six $\sigma$-bonds required for octahedral shape coordination to create a well-defined quintuple layer.  However, the Se1 atom (the Se atom near the van der Waals gap or surface) has three electrons that remain unpaired after hybridization.  To reduce Coulomb repulsion, the maximum separation between the three unpaired electrons on the Se1 atom corresponds to a planar configuration with 120$^\circ$ angles between the electrons, i.e., a trimer, where three out of the six neighboring Se1 atoms in the same layer can participate in three $\pi$-bonds as shown in the inset of Fig.~\ref{fig:fig-hybridBi2Se3}.  We will refer to the three symmetrically distributed $\pi$-bonds as a $\pi$-bond trimer, and a $\pi$-electron is used to denote the unpaired electron resulting from the breaking of a $\pi$-bond.

The $\pi$-bond trimer on the Se1 surface of Bi$_2$Se$_3$ has two equivalent states within a six-fold symmetry coordination, as illustrated in the inset of Fig.~\ref{fig:fig-hybridBi2Se3}.  Because the $\pi$-bond is actually formed by an extremely weak instantaneous attractive force due to minimum orbital overlap,\cite{textbook} entropy is maximized by breaking and re-forming the $\pi$-bonds to access the true ground state that minimizes the Gibbs free energy $\Delta G=\Delta \textbf{H}-T\Delta S$.  Clearly, the two equivalent states of the $\pi$-bond trimer in space and time correspond to a conjugated system of two quantum mechanical ensembles without a stability preference.\cite{Shu2012}   A local energy exchange mechanism between the $\pi$-bond trimer conjugated system and six-way Se-Se connection on the Bi$_2$Se$_3$ surface (inset of  Fig.~\ref{fig:fig-hybridBi2Se3}) has been proposed to rationalize the unique Bi$_2$Se$_3$ surface conduction mechanism analogous to a dynamic three-way cross-bridge connecting six ports.  The proposed surface conduction mechanism for graphene and Bi$_2$Se$_3$ surfaces is drastically different from the traditional view of metallic conduction that is based on the diffusion of itinerant electrons in the conduction band.\cite{Shu2012}

\section{\label{sec:level1} $\alpha-Sn$: $\pi$-bond dimers in four-fold symmetry coordination\protect\\}

\begin{figure}
\includegraphics[width=3.5in]{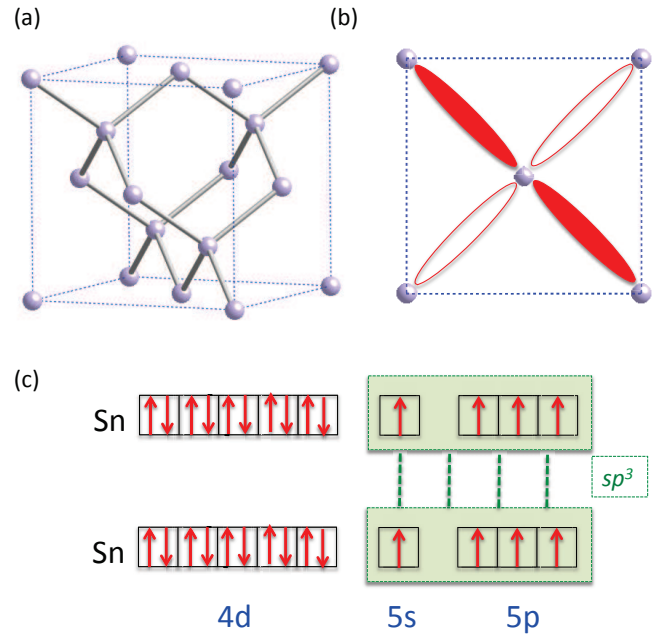}
\caption{\label{fig:fig-alphaSn}(color online) (a) The diamond crystal structure of $\alpha$-Sn, where all Sn atoms are in identical tetrahedral coordination.  (b) The (001) surface obtained by breaking two of the four $\sigma$-bonds, creating two $\pi$-electrons per surface Sn, is shown.  The two $\pi$-electrons are 180$^\circ$ apart and form a $\pi$-bond dimer with two identical configurations, i.e., one conjugated system is formed for two equivalent states.  c) The hybrid orbital model for $\alpha$-Sn is shown, where $sp^3$ hybridization is satisfied for Sn ([Kr]$4d^{10}5s^25p^2$) with four neighboring Sn atoms in tetrahedral coordination. }  
\vspace{-5mm}
\end{figure}

Topological insulators with a four-fold symmetry in 2D band projection have also been predicted and subsequently found, such as the strained $\alpha$-Sn and HgTe/CdTe quantum well with a zinc-blende structure.\cite{Fu2007, Dai2008}   Fig.~\ref{fig:fig-alphaSn} shows the diamond structure of $\alpha$-Sn.  The tetrahedral coordination for each Sn atom ([Kr]$4d^{10}5s^25p^2$) requires $sp^3$ hybridization using the four outer shell electrons in both the $5s$ and $5p$ orbitals.  A (001) surface cut breaks two of the four $\sigma$-bonds, leaving two unpaired $\pi$-electrons on the surface that can form two $\pi$-bonds with two of the four neighboring Sn atoms on the same (001) surface, as shown in Fig.~\ref{fig:fig-alphaSn}(b). To avoid Coulomb repulsion, the maximum separation between the $\pi$-bonds corresponds to an angle of 180$^\circ$ between the bonds, i.e., a $\pi$-bond dimer. There are two equivalent states for each $\pi$-bond dimer in space and time.  Because the $\pi$-bond is extremely weak, being induced by transient attractions through minimal orbital overlap, it can be easily perturbed by an entropy increase to lower the Gibbs free energy and access the true ground state.  We can thus identify the $\pi$-bond dimer in four-fold symmetry coordination on the surface as a conjugated system, in a similar manner as for the $\pi$-bond trimer in six-fold symmetry coordination on the Bi$_2$Se$_3$ (111)$_r$ surface discussed above.\cite{Shu2012}  The same cross-bridge model for surface conduction could also be applied to the strained $\alpha$-Sn as a topological insulator when the $\pi$-bond dimer is semi-localized to maintain a dynamic conjugated system without preferrential localization or itinerancy.  

\section{\label{sec:level1} Parity symmetry breaking of $\pi$-bonds in 2D \protect\\}

\begin{figure}
\includegraphics[width=3.5in]{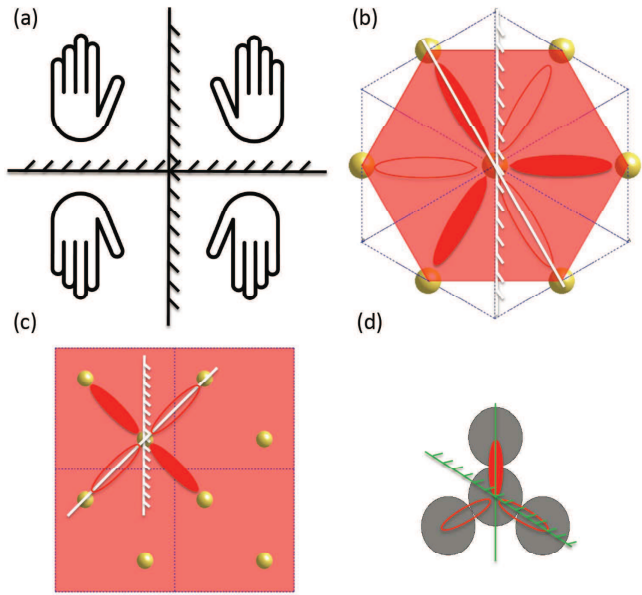}
\caption{\label{fig:fig-Parity}(color online) (a) Parity and handedness for chirality is shown in 2D. (b) A $\pi$-bond trimer in a six-fold symmetry coordination is shown.  (c) A $\pi$-bond dimer in a four-fold symmetry coordination is shown. (d) A lone $\pi$-bond in a three-fold symmetry coordination is shown.}  
\vspace{-5mm}
\end{figure}

To date two main relevant types of high symmetry surface atomic arrangements have been found in the topological insulators with either six- or four-fold symmetry.  The first type is found in the (001)$_h$ plane of layered Bi$_2$Se$_3$, Bi$_{1-x}$Sb$_x$, and in the (111) plane of a rocksalt structure, such as SnTe,\cite{Zhang2009, Hsieh2012} while the second type is found in the (001) plane of zinc-blende structures, such as strained HgTe and $\alpha$-Sn.\cite{Fu2007, Dai2008}   We can identify 2D parity symmetry for both a $\pi$-bond trimer in six-fold symmetry coordination and a $\pi$-bond dimer in four-fold symmetry coordination, as illustrated in Fig.~\ref{fig:fig-Parity}.  While parity transformation is related to the inversion of the spatial coordinates, the x- and y-axes cannot be inverted simultaneously to obtain an inverted image in 2D, i.e., the determinant of the parity operation \textbf{P} must be -1, as illustrated in Fig.~\ref{fig:fig-Parity}(a). If this condition is not satisfied, flipping the signs of both the x- and y-axes only produces a 180$^\circ$ rotation without mirror inversion.  The parity transformation can test for chirality by determining whether a mirror image is obtained following the parity operation.  Chirality is confirmed by the breaking of parity symmetry.  We can select between the two high-symmetry states by rejecting a mirror line that results in identity transformation such that parity symmetry is broken for surfaces with $\pi$-bond-trimers in six-fold symmetry coordination and $\pi$-bond dimers in four-fold symmetry coordination, as illustrated by the mirror lines in Fig.~\ref{fig:fig-Parity}(b)-(c).  In fact, similar parity symmetry breaking is also observed in the honeycomb graphene structure as shown in Fig.~\ref{fig:fig-Parity}(d), for a properly chosen mirror line.\cite{Shu2012}  We submit that 2D parity symmetry breaking is consistent with the unique character of surface chiral electrons with E(k) dispersion in a Dirac cone shape for a topological insulator.  

\section{\label{sec:level1} surface spin polarization and preferred chirality\protect\\ }

The existence of chiral fermions in topological insulators in edge state (in 2D) and surface state (in 3D) with spins normal to the direction of their lattice momentum has been predicted by the quantum spin Hall effect.\cite{Hasan2010}   The locked spin momentum and $\pi$ Berry's phase in Bi$_2$Se$_3$ is regarded as one of the defining properties of topological insulators, as confirmed by spin-ARPES.\cite{Hsieh2009}  Chirality for a $\pi$-bond trimer in six-fold symmetry coordination and a $\pi$-bond dimer in four-fold symmetry coordination must be closely connected with the surface spin polarization.  Generally, right-handed (RH) or left-handed (LH) chirality is expected to be statistically equal, with no preference based on entropic considerations.  However, the discovery of surface spin polarization strongly suggests that RH and LH chirality is not balanced for $\pi$-bond trimers and dimers on a topological insulator surface.  

We submit that a novel surface spin-orbit coupling results from the chirality imbalance that follows 2D parity symmetry breaking such that the $\pi$-electrons effectively behave as if they were in a non-zero magnetic field generated by the uncanceled orbital angular momentum, analogous to classical mechanics.   This subtle effect must be a quantum phenomenon arising from entropic considerations for a dynamic conjugated system of two energetically equivalent ensembles of different $\pi$-bond arrangements.  Nature clearly provides an even subtler mechanism for the surface $\pi$-electrons to access the true ground state through an additional magnetic energy gain ($-H\Delta M$) in the Gibbs free energy of $\Delta G=\Delta\textbf{H}-T\Delta S-H\Delta M$, in addition to the entropic contribution of $-T\Delta S$.    

Magnetic impurities, such as Mn, have been shown to be capable of opening the gap in the Dirac cone for Bi$_2$Se$_3$,\cite{Hsieh2009} which can also be explained intuitively using our proposed $\pi$-bond model.  Mn doping effectively destroys 2D parity symmetry by adding a strong local magnetic field to polarize the electron spins, creating a strong and random handedness preference.  In other words, the system initially requires a very small gain in the effective magnetic energy $-H\Delta M$ to compete with the entropy increase $-T\Delta S$ from the slight chirality imbalance, however, the strong local field created by the large Mn magnetic moment breaks the subtle imbalance and shifts the system toward a new ground state within the correlation length of the magnetic moment, which does not allow for the formation of a dynamic $\pi$-bond conjugated system when the handedness is biased beyond being balanced by the entropy.

\section{\label{sec:level1} $\pi$-bond model predictions for topological insulators \protect\\}

\begin{figure}
\includegraphics[width=3.5in]{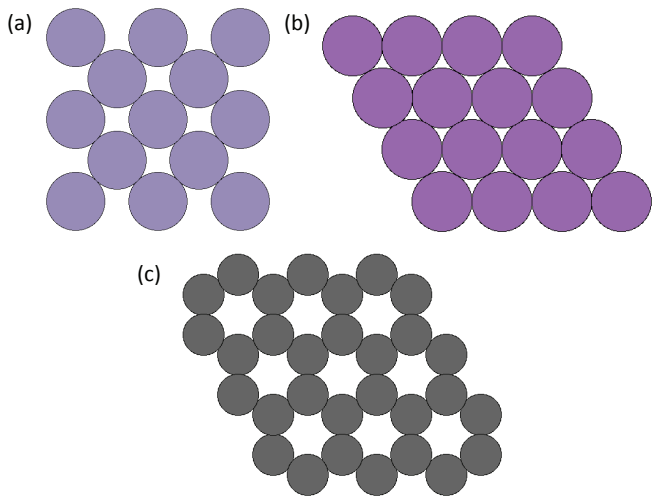}
\caption{\label{fig:fig-2Dpacking}(color online) Three typical surface atomic packings with various coordination numbers: (a) four-fold square packing, (b) six-fold hexagonal packing, and  (c) three-fold honeycomb packing.}  
\vspace{-5mm}
\end{figure}

A general rule in crystal formation is that the number of bonds is maximized for the greatest crystal energy gain.  Surface atomic packings typically fall into one of the three categories almost exclusively as a result of atomic size matching and outer shell electron bonding in 3D, as illustrated in Fig.~\ref{fig:fig-2Dpacking}.   The existence of $\pi$-bond trimers in six-fold symmetry coordination and $\pi$-bond dimmers in a four-fold symmetry coordination is a commonly found feature in topological insulators.  The honeycomb structure that has been predicted for compounds such as LiAuSe could be closely related to a $\pi$-bond conjugated system in three-fold symmetry coordination.\cite{Zhang2011}  A typical signature of $\pi$-bonds on the surfaces of these three classes of topological insulators is 2D parity symmetry breaking, as illustrated in Fig.~\ref{fig:fig-Parity}.  In fact, we submit that 2D parity symmetry breaking in a $\pi$-bond conjugated system is the only criteria by which a material may be considered to be topologically trivial or nontrivial.   

The necessary condition for the existence of a dynamic $\pi$-bond conjugated system in topological insulators is the requirement of weak covalent bonds in the material.  For example, nearly all of the verified topological insulators fall into the category of narrow band gap semiconductors, the main character of which is a low melting point resulting from weak covalent bonding.  If the material is a semimetal with a small band overlap, the band gap can be adjusted through proper atom substitution, such as tuning Bi into a narrow band gap semiconductor so that the semiconductor acts as topological insulator Bi$_{1-x}$Sb$_x$.\cite{Hsieh2008}  In contrast, for insulators with covalent bonds that have a slight polarity to ionic character due to a relatively large EN difference, atomic substitution can narrow the band gap by EN adjustment; for example, the EN difference of PbTe with a rocksalt structure (EN of Pb=2.33 and EN of Te=2.1) may be tuned with Sn substitution (EN of Sn=1.96).\cite{Xu2012}  We conclude that the role of the strain in tuning a topological state is actually a side effect of the atomic substitution because the strain is introduced by the unavoidable mismatch in atomic sizes through substitution, although the goal is electronic band gap size tuning.  

Based on our proposed $\pi$-bond model, HgTe with a zinc-blende structure is likely to be topologically nontrivial due to a low melting point from weak covalent bonding (EN of Hg=2.0 and Te=2.1), and the existence of surface $\pi$-bond dimers in four-fold symmetry coordination on (001) planes,  in agreement with the prediction by Chadov \textit{et al.}\cite{Chadov2012}   However, we expect AmN with a rocksalt structure to be topologically trivial,\cite{Zhang2012} mainly due to the particularly large EN difference between the Am and N atoms.  The existence of the dynamic $\pi$-bond conjugated system on topological insulator surfaces requires a low surface energy to accommodate the entropic factor on lowering the Gibbs free energy.  The energies of the itinerant electrons in metals and the unpaired electrons resulting from the strong covalent bonds in a wide band gap insulator being broken are clearly too high to maintain even $\pi$-bonds on the surface.

\section{\label{sec:level1}Conclusions\protect\\}

In summary, we have applied a hybrid orbital model to typical topological insulators for Bi$_2$Se$_3$ and $\alpha$-Sn based on crystal symmetry, surface atomic coordination, and the outer-shell electrons.  The identification of 2D parity symmetry breaking leads to a reasonable explanation for the chiral character of surface electrons, which is implied by the quantum spin Hall effect, and this chirality imbalanced can be used to interpret the spin polarization found in the surface band of a Dirac cone.  It is clear that the $\pi$-bond is key to understanding the surface electronics, as it has been repeatedly found in the ideal 2D graphene electron gas and topological insulator surfaces with similar Dirac cones.

\section*{Acknowledgment}
We thank G. Y. Guo for helpful discussions.  FCC acknowledges the support provided by NSC-Taiwan under project number  NSC 101-2119-M-002-007.  GJS acknowledges the support provided by NSC-Taiwan under project number NSC 100-2112-M-002-001-MY3.


\begin{thebibliography}{99}

\bibitem{Hasan2010} M. Z. Hasan and C. L. Kane, Rev. Mod. Phys. \textbf{82}, 3045 (2010).
\bibitem{Huang2012} F.-T. Huang, M.-W. Chu, H. H. Kung, W. L. Lee, R. Sankar, S.-C. Liou, K. K. Wu, Y. K. Kuo, and F. C. Chou, Phys. Rev. B, \textbf{86}, 081104 (2012).
\bibitem{Qi2011} X.-L. Qi and S.-C. Zhang, Rev. Mod. Phys. \textbf{83}, 1057 (2011).
\bibitem{Hsieh2009} D. Hsieh, Y. Xia, D. Qian, L. Wray, J. H. Dil, F. Meier, J. Osterwalder, L. Patthey, J. G. Checkelsky, N. P. Ong, A. V. Fedorov, H. Lin, A. Bansil, D. Grauer, Y. S. Hor, R. J. Cava, and M. Z. Hasan, Nature \textbf{460}, 1101 (2009).
\bibitem{Kane2006} C. L. Kane and E. J. Mele, Science, \textbf{314},1692 (2006).
\bibitem{Zhang2009} H. Zhang, C. X. Liu, X. Qi, X. Dai, Z. Fang, and S. C. Zhang, Nat. Phys. \textbf{5}, 438 (2009).
\bibitem{Shu2012} G. J. Shu and F. C. Chou,  arXiv:1212.5982.
\bibitem{textbook} L. G. Jr. Wade, Organic Chemistry (7th Edition), Prentice Hall 2009.
\bibitem{Hsieh2008} D. Hsieh, D. Qian, L. Wray, Y. Q. Xia, Y. S. Hor, and R. J. Cava, Nature \textbf{452}, 970 (2008).
\bibitem{Hsieh2012} Timothy H. Hsieh, Hsin Lin, Junwei Liu, Wenhui Duan, Arun Bansil, and Liang Fu, Nat. Comm. \textbf{3}, 982 (2012)
\bibitem{Tanaka2012} Y. Tanaka, Zhi Ren, T. Sato, K. Nakayama, S. Souma, T. Takahashi, Kouji Segawa, and Yoichi Ando, Nature Physics 8, 800Ð803 (2012)
\bibitem{Fu2007} L. Fu, C. L. Kane, Phys. Rev. B \textbf{76}, 045302 (2007).
\bibitem{Dai2008} X. Dai, T. L. Hughes, X. L. Qi, Z. Fang, and S. C. Zhang, Phys. Rev. B \textbf{77}, 125319 (2008).
\bibitem{Katsnelson2006} M. I. Katsnelson, K. S. Novoselov, and A. K. Geim, Nature Physics \textbf{2}, 620 (2006).
\bibitem{Zhang2011} H. J. Zhang, S. Chadov, L. MŸchler, B. Yan, X. L. Qi, J. KŸbler, S. C. Zhang, and C, Felser, Phys. Rev. Lett., \textbf{106}, 156402 (2011).
\bibitem{Xu2012} S.-Y. Xu, C. Liu, N. Alidoust, M. Neupane, D. Qian, I. Belopolski, J. D. Denlinger, Y.-J. Wang, H. Lin, L. A. Wray, G. Landolt, B. Slomski, J. H. Dil, A. Marcinkova, E. Morosan, Q. Gibson, R. Sankar, F. C. Chou, R. J. Cava, A. Bansil, and M. Z. Hasan, Nat. Comm. \textbf{3}, 1192 (2012).
\bibitem{Chadov2012} S. Chadov, J. Kiss, C. Felser, K. Chadova, D. Kšdderitzsch, J. Min‡r, and H. Ebert, arXiv:1207.3463 (2012).
\bibitem{Zhang2012} X. Zhang, H. Zhang, J. Wang, C. Felser, S. C. Zhang, Science \textbf{335}, 1464 (2012)

\end{thebibliography}

\pagebreak

\end{document}